# Externally triggered coherent two-photon emission from hydrogen molecules

Yuki Miyamoto[1,*], Hideaki Hara[1], Takahiko Masuda[1], Noboru Sasao[1], Minoru Tanaka[2], Satoshi Uetake[3], Akihiro Yoshimi[1], Koji Yoshimura[1], and Motohiko Yoshimura[3]

[1]*Research Core for Extreme Quantum World, Okayama University, Okayama 700-8530, Japan*

[2]*Department of Physics, Osaka University, Toyonaka, Osaka 560-0043, Japan*

[3]*Research Center of Quantum Universe, Okayama University, Okayama 700-8530, Japan*

*E-mail: miyamo-y@okayama-u.ac.jp

**Abstract**

We report coherent enhancement of two-photon emission from the excited vibrational state of molecular hydrogen triggered by irradiating mid-infrared pulses externally. We previously observed the two-photon emission triggered by the internally generated fourth Stokes photons. By injecting independent mid-infrared pulses externally, it is possible to control experimental parameters and investigate the mechanism in more detail. In this article, we describe the two-photon emission using the external trigger pulses. Its spectrum and dependence on the energy and timing of the trigger pulse are presented along with numerical simulations based on the Maxwell-Bloch equations. The measured number of emitted photons is $6\times10^{11}$ photons/pulse and the resulting enhancement factor from the spontaneous emission is more than $10^{18}$. This value is three orders of magnitude higher than that of the previous experiment. External control of emission process is expected to be essential for observation of weaker process of radiative emission of neutrino pair.

*1. Introduction* Coherence of matter interacting with photons is one of the key concepts in atomic, molecular and optical physics. This many-body effect makes phenomena quite different from cases of ensemble of incoherent particles in which destructive and random interference works. A variety of coherent transient effects, such as free induction decay [1], spin or photon echo [2], and super-radiance [3], have been investigated for the last several decades. Some of present authors have proposed to study rare emission of plural particles involving neutrinos (radiative emission of neutrino pair: RENP) by coherent amplification mechanism [4]. The ultimate goal of the proposal is to investigate unknown neutrino properties such as their absolute masses, mass type (Dirac or Majorana), and CP-violating phases by measuring the energy spectrum of RENP.

We recently succeeded in observing the "internally triggered" two-photon emission from gaseous para-hydrogen (p-$H_2$) [5]. In the experiment, coherence of the p-$H_2$ vibration was prepared by adiabatic Raman process [6]. When the coherence is high enough, higher order Raman scattering occurs and generates more than ten sidebands from UV to mid-infrared. The observed two-photon emission was triggered by the mid-infrared photons which were "internally" generated as a fourth Stokes Raman sideband. The observation of the two-photon emission assured that coherence can accelerate rare emission processes of plural particles (two photons, in this case).

**Table 1** Relevant photons and their wavelength

| symbols | wavelength (nm) | description |
|---|---|---|
| $\omega_0$ | 532.2 | Driving Field |
| $\omega_{-1}$ | 683.6 | Driving Field (first Stokes) |
| $\omega_{-4}$ | 4662 | Fourth Stokes (internal trigger) |
| $\omega_{-5}$ | 4959 | Two-photon emission partner triggered by $\omega_{-4}$ |
| $\omega^e_0$ | 4587 | External trigger pulse |
| $\omega^e_{-1}$ | 5048 | Two-photon emission partner triggered by $\omega^e_0$ |

While use of the fourth Raman sideband as a trigger was an easy way, we could not control trigger parameters independently, such as energy and timing of the trigger. In the present work, we prepared a mid-infrared laser as a triggering source and measured dependences of the two-photon emission rate on the trigger energy and timing to investigate the enhancement mechanism in more detail. The results are compared with numerical simulations based on the Maxwell-Bloch equations. External triggering is one of key techniques for RENP because RENP should be induced by triggering photons and spectra of RENP might be measured by scanning the wavelength of the trigger. Demonstration of external control is, therefore, an important step to the observation of RENP.

The rest of the paper is organized as follows. In the next section, we describe experimental setup. In Sect. 3, the experimental results are presented and compared with the simulation. Conclusions are given in Sect. 4.

*2. Experimental Setup* Energy diagram and experimental setup of the current study are shown in Fig. 1. They are basically same as those of the previous paper [5]. Here, we describe current setup briefly, and mention the difference from the previous one. Table 1 summarizes the wavelengths of photons having important roles in the present study.

Para-hydrogen gas was prepared by passing normal hydrogen gas into a magnetic catalyst, Fe(OH)O, cooled to about 14 K by a Gifford-McMahon refrigerator. Residual ortho-hydrogen concentration was estimated to be less than 500 ppm. The prepared gas was transferred into a copper cell with 2 cm in diameter and 15 cm in length. Both sides of the cell were sealed with Magnesium Fluoride ($MgF_2$) windows of 5 mm thickness. The temperature of the cell was kept at about 78 K in a liquid-nitrogen cryostat whose optical windows were also made of $MgF_2$. All experiments discussed below were performed with 60 kPa p-$H_2$ target.

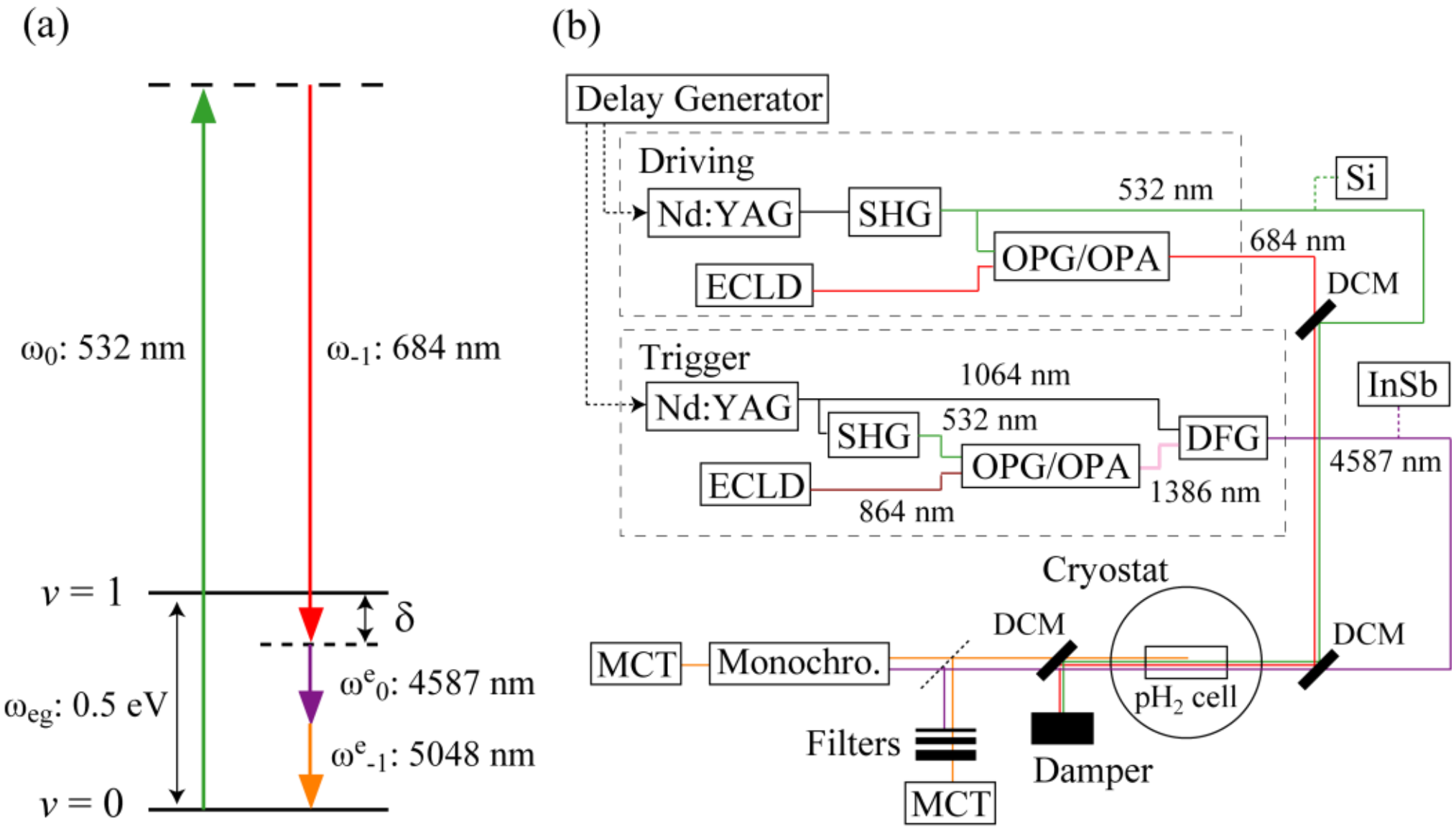

Figure 1: Energy diagram (a) and experimental setup (b). DCM: Dichroic Mirror, DFG: Difference Frequency Gereration, ECLD: External Cavity Laser Diode, InSb: Indium Antimony photo-detector, MCT: Mercury Cadmium Tellurium photo-detector, Monochro.: Monochromator, OPG: Optical Prametric Generator, OPA: Optical Parametric Amplification, SHG: Second Harmonic Generation, Si: Silicon photo-detector

Coherence of vibrational states of the target was provided via adiabatic Raman process driven by two visible laser pulses: 532 nm ($\omega_0$) and 684 nm ($\omega_{-1}$) [6], which are called driving fields in this article. This process and experimental setup are identical to those of the previous experiment. The detail is described in the previous paper [5]. Typical pulse energy of each driving laser was 5 mJ/pulse. Duration (full width at half maximum) of each pulse was about 9 ns and 6 ns, respectively. Energy difference of the two photons is equal to the vibrational energy of p-$H_2$, $\omega_{eg}$ (4161 $cm^{-1}$ ~ 0.5 eV) [7], except for small detuning $\delta = \omega_{eg} - (\omega_0 - \omega_{-1})$. The δ can be varied up to ± 1 GHz by changing wavelength of 684-nm pulses. In the present experiment, however, the detuning δ was fixed to be −160 MHz during the present experiments where the intensity of the two-photon emission is maximum. The uncertainty of the detuning was ±75 MHz due to the absolute accuracy in the wavelength meter (HighFinesse WS-7). By comparing energies of Raman sidebands and those of the Maxwell-Bloch numerical simulation, prepared coherence is estimated to be about 0.04 (the maximum coherence is 0.5), which is almost same as that of the previous paper. Decoherence used in the simulation was taken from observed Raman linewidth (65 MHz in a half width at half maximum) [8]. Furthermore, intensities of the driving fields were adjusted to represent the observed Raman sidebands energy ratio. The determined effective intensity of 532-nm and 684-nm field was 340 and 190 MW/$cm^2$, respectively, that is smaller than actual intensities.

The most important modification from the previous experiment is the injection of the third laser pulses at 4587 nm as the external trigger. Its duration was approximately 2 ns (FWHM) and typical energy was 150 μJ/pulse. The external trigger pulses were prepared in three steps described below. In the first step, a periodically poled lithium niobate (PPLN, Conversion, MSFG1-20) was pumped by second harmonics (532 nm) of a Nd:YAG laser (Continuum, Surelite I) with injection of a continuous-wave diode laser at 864 nm. The pumped PPLN emitted 864-nm and 1386-nm pulses by optical parametric generation. The 864-nm and 1386-nm pulses were amplified in the second step by optical parametric amplification using Lithium triborate (LBO, Tecrys Ltd.) crystals pumped by the 532-nm pulses. In the third steps, 1064-nm pulses from the identical Nd:YAG laser and the amplified 1386-nm pulses were collinearly introduced into Potassium Titanyle Arsenate (KTA, Crystech) crystals and 4587-nm pulses were generated via difference frequency generation.

The timing of pulse peaks of two driving lasers coincided with each other less than 50 ps difference by adjusting optical path length. The timing jitter between them was negligible because these two driving pulses were originated from the same pulse. We assumed that the two pulses arrived simultaneously at the target. A delay generator (Stanford research systems, DG645) synchronized the two Nd:YAG lasers and controled the mutual timing between the driving fields and the external trigger pulses. The jitter of the mutual timing was estimated to be about 0.8 ns (standard deviation). The mutual timing was monitored shot-by-shot by measuring a pulse shape of a small portion of one of the driving lasers (532 nm) and the trigger pulses using a Silicon (Si) photo-diode (Hamamatsu, S5973) and an Indium-Antimony (InSb) detector (Hamamatsu, P5968-100), respectively. The timing resolution was approximately 0.1 ns and the absolute accuracy of the mutual timing was estimated to be less than 0.5 ns.

The three pulses (532, 684 and 4587 nm) were aligned collinearly by two custom-made dichroic mirrors and irradiated the target. All three pulses were horizontally polarized. They were operated at a repetition rate of 10 Hz. If the coherence of p-$H_2$ is high enough, the enhanced two-photon emission occurs and photons at 5048 nm are emitted. The intense 532 and 684-nm photons were eliminated from the output pulses by another dichroic mirror put just behind the cryostat. Transmitted pulses still contained the several frequency components due to the higher order Raman scattering. Band-pass and/or long wavelength pass filters, therefore, were used to detect two-photon emission signals. A monochromator (Princeton Instruments, Acton SP2300) was also used to measure the spectra of the emitted pulses. A Mercury-Cadmium-Tellurium (MCT) detector was used to detect mid-infrared pulses (Vigo system, PV-2TE-6). Neutral density filters were also used to prevent saturation of the detectors.

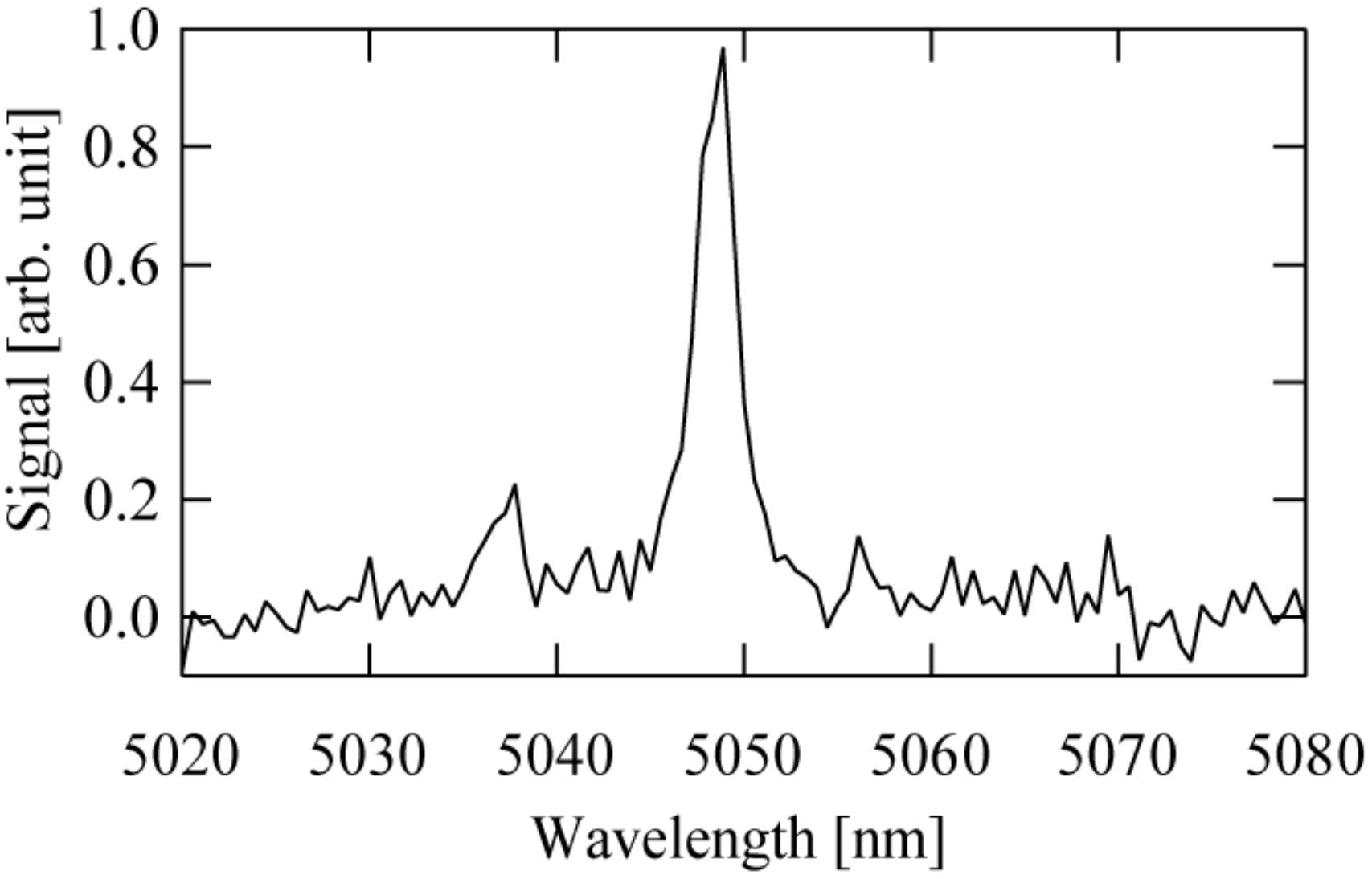

Figure 2: Observed spectrum of externally triggered two-photon emission.

*3. Results and Discussion* Figure 2 shows observed spectrum of the two-photon emission measured by the monochromator. The observed center frequency was 5048 nm and consistent with the expectation from the energy conservation (that is, $\omega_0 - \omega_{-1} = \omega^e_0 + \omega^e_{-1}$) within 1 nm. Linewidth was attributed to resolution of the monochromator. We measured the pulse energy of the signal through band-pass filters (Thorlabs, FB5250-500) instead of the monochromator because the transmittance of the filter was able to be estimated more easily than that of the monochromator. The MCT detector's responsivity was also evaluated using the trigger laser at 4587 nm with the assumption that responsivities at 5048 nm ($\omega^e_{-1}$) and 4587 nm ($\omega^e_0$) were same. Considering the optical transmittance from the p-$H_2$ target cell to the detector, the number of photons was estimated to be $6\times10^{11}$ photons/pulse in the p-$H_2$ target. We have defined "enhancement factor" as a ratio of the observed photon number to that expected due to spontaneous two-photon emissions with experimental acceptance [5]. The enhancement factor in the current experiment was found to be more than $10^{18}$ and three orders of magnitude larger than that in the previous experiment, where the two-photon emission was triggered internally by the fourth Stokes [5]. The higher intensity of the external trigger pulse than that of the fourth Stokes should contribute this improvement. Energy ratio of the two photon emission partner at 5048 nm ($\omega^e_{-1}$) to the external trigger at 4587 nm ($\omega^e_0$) was ~$10^{-4}$, which was same order as that of the two photon emission partner at 4959 nm ($\omega_{-5}$) to the internal trigger light at 4662 nm ($\omega_{-4}$). As discussed later, output energy of the two-photon emission is expected to be proportional to intensity of the trigger. The observed energy ratios also support this expectation.

Divergence of the 5048-nm beam ($\omega^e_{-1}$) was roughly estimated by a home-made slit-scan profiler. The divergence half angle was found to be less than 10 mrad, which is consistent with diffraction-limited divergence.

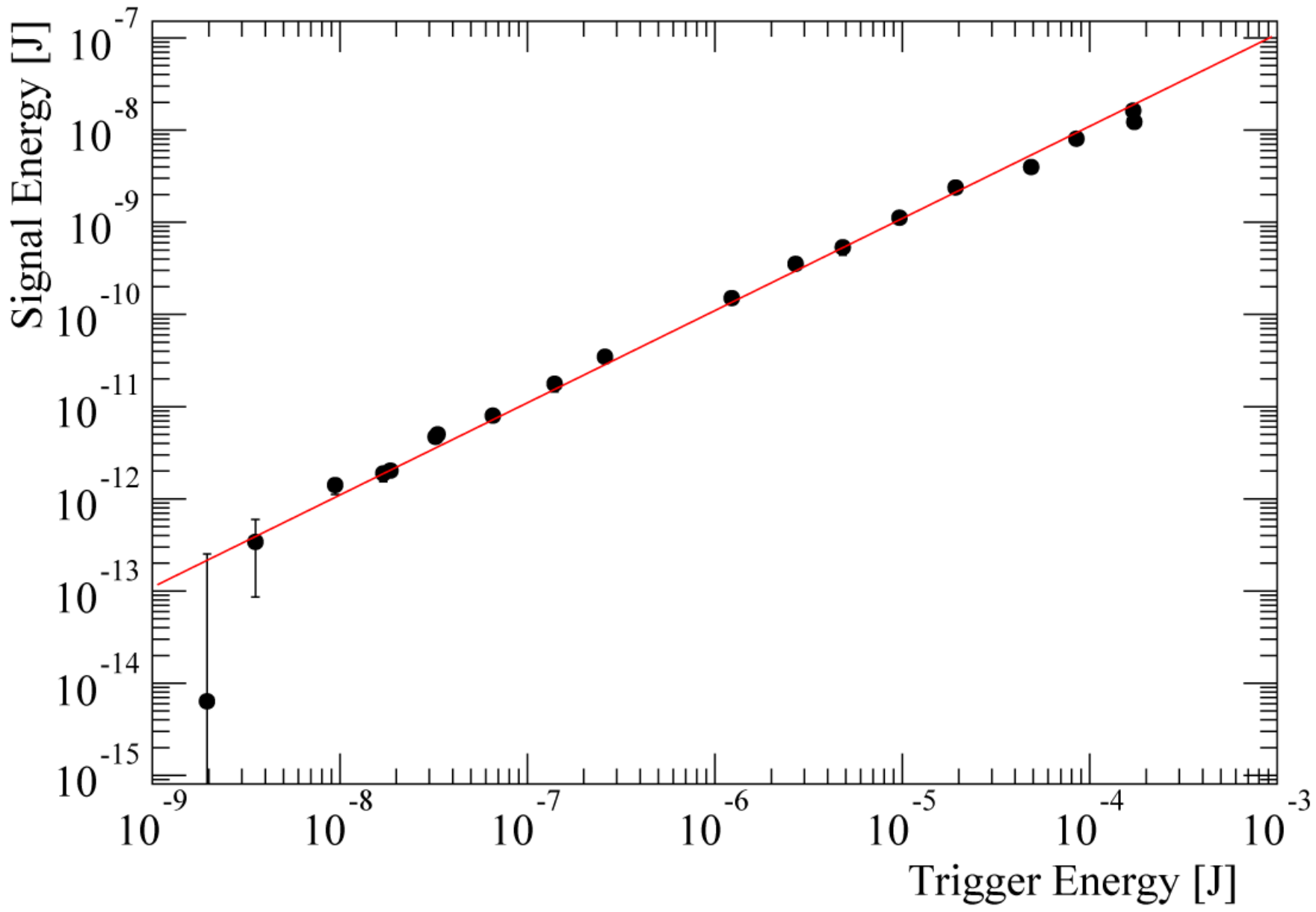

Figure 3: Dependence of the signal energy on the trigger energy.

We also observed first anti-Stokes light of the external trigger at 1577 nm as expected. The intensity ratio of the anti-Stokes to the two-photon emission partner was about 50, whose order was consistent with the numerical simulations. The higher order Raman scattering of the external trigger could not be observed in the present experiment.

Figure 3 shows the dependence of the energy of the two-photon emission partner on the trigger energy. The trigger energy was reduced by inserting neutral density filters. The observed dependence can be represented by the linear function with a zero intercept over more than four orders of magnitude. The Maxwell-Bloch equations also expected the linear behavior in the limit of weak trigger intensity. The absence of the observable higher-order Raman scattering of the trigger also suggests that the trigger energy is low enough. The energy of trigger laser may be distributed to higher order anti-Stokes photons and the two-photon emission energy is expected to deviate from linear dependence when the trigger energy becomes comparable to those of the driving fields.

Dependence of the two-photon emission rate on the mutual timing between the trigger pulses and the driving pulses is shown in Fig. 4. The timing of each shot was obtained from the pulse shapes of the Si and InSb photo-detectors and was binned in 0.5 ns. Positive values in the horizontal axis mean that the trigger pulses arrive at the target after the driving pulses. The observed dependence showed asymmetric behavior with respect to the origin. A solid line in Fig. 4 shows a corresponding simulation result. It should be mentioned that the peak height of both traces is normalized to be unity. The decoherence time taken from the Raman linewidth [8] was again used. The effective intensities of the driving fields were fixed to the values determined by the simulation of the Raman sidebands described in Sect. 2. The simulation could qualitatively reproduce the experimental result without adjusting parameters while the simulation result showed a longer tail at the positive side and a small bump at the negative side. In the simulation, we have considered one position coordinate along the light propagation and ignored the other two spatial coordinates. The observed “delayed peak” can be explained qualitatively by development and decay of coherence in a simple optical Bloch picture. In the current experiment, two-photon Rabi frequency is so small (less than 10 MHz) that the period of Rabi oscillation is longer than the driving pulse width. It indicates that the driving fields pass away before a single Rabi oscillation. Coherence, therefore, reaches its maximum after the peak of driving fields and then decays due to low driving intensity and decoherence. The two-photon emission reflects the coherence development and shows the delayed peak. The peak position is determined by the Rabi frequency, the driving field duration and the decoherence. This kind of the “delayed peak” was also reported in a stimulated Raman process [9].

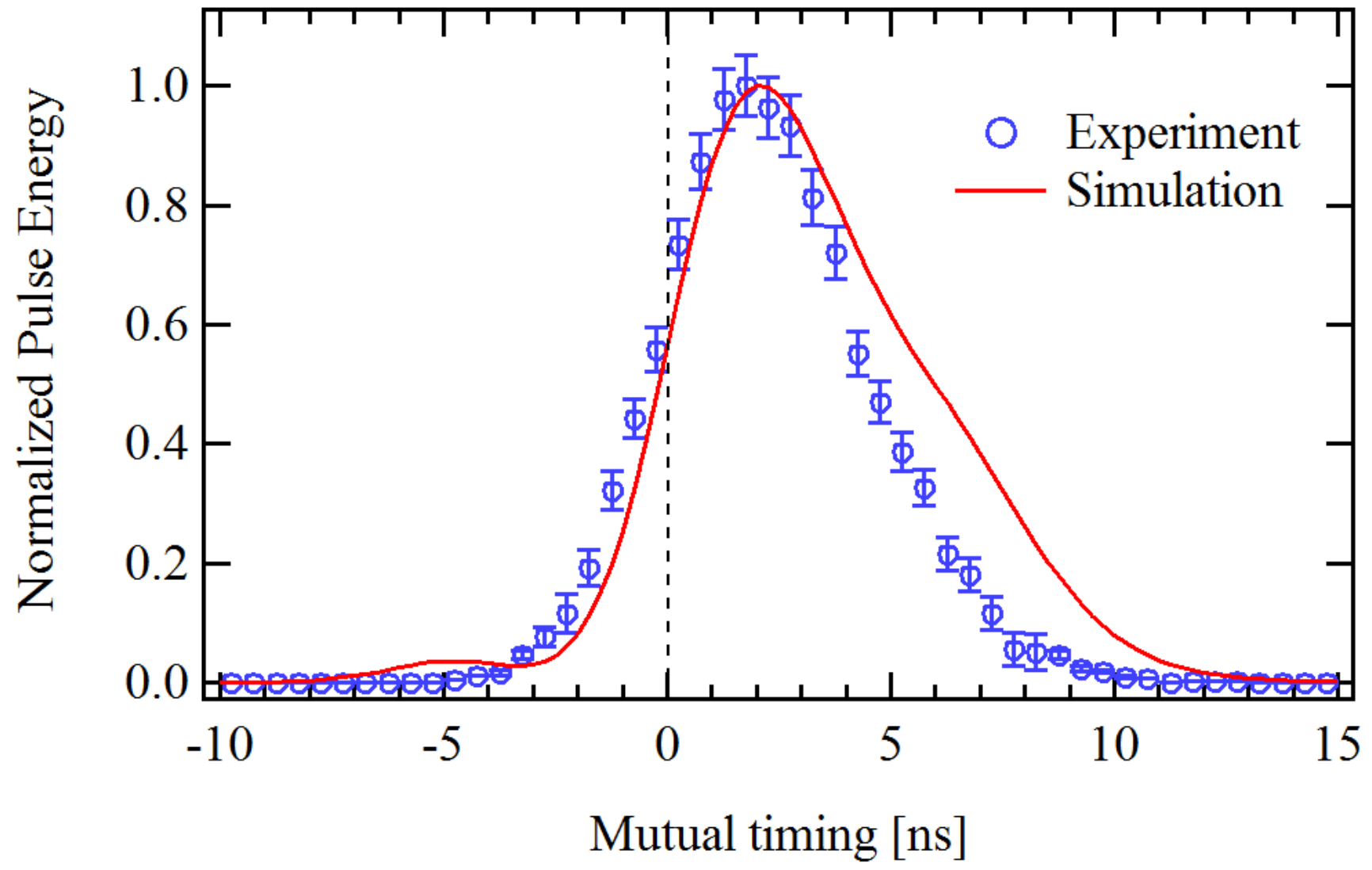

Figure 4: Observed dependence of two-photon emission signal on mutual timing between the driving pulses and the triggering pulses (blue open circles) and corresponding result of the numerical simulation (a red solid line). Positive values in the horizontal axis mean that the trigger pulses arrive at the target after the driving pulses.

*4. Conclusions* In this work, we have observed coherent enhancement of the two-photon emission from the vibrational state of p-$H_2$ with the method of triggering by externally injected pulses. The enhancement from its spontaneous rate has been three orders of magnitude larger than that of the previous observation triggered by the internally generated Stokes light.

Dependences of the emission rate on the trigger energy and timing were measured. The rate is proportional to the trigger energy and enhancement is asymmetric with respect to the origin of mutual timing. These behaviors agree qualitatively with the numerical simulation based on the Maxwell-Bloch equations, indicating fairness of our formulation of the coherent enhancement of two-photon emission.

**Acknowledgement**

We thank Professor J. Tang for providing us mid-infrared detectors. This research was partially supported by Grant in-Aid for Scientific Research on Innovative Areas "Extreme quantum world opened up by atoms" (21104002), Grant-in-Aid for Scientific Research A (21244032), Grant-in-Aid for Scientific Research C (25400257), Grant-in-Aid for Challenging Exploratory Research (24654132), Grant-in-Aid for Young Scientists B (25820144), and Grant-in-Aid for Research Activity start-up (26887026) from the Ministry of Education, Culture, Sports, Science, and Technology.